\documentclass[12pt,preprint]{aastex}

\shorttitle{The Horizontal Branch of NGC~1851}
\shortauthors{Salaris et al.}

\begin{document}
\title{The Horizontal Branch of NGC~1851: constraints on the cluster subpopulations}

\author{M. Salaris} 
\affil{Astrophysics Research Institute, Liverpool John Moores University,
Twelve Quays House, Egerton Wharf, Birkenhead CH41 1LD, UK}
\email{ms@astro.livjm.ac.uk}

\author{S. Cassisi and A. Pietrinferni}
\affil{INAF - Osservatorio Astronomico di Collurania, Via M.\ Maggini,
I-64100 Teramo, Italy}
\email{cassisi,pietrinferni@oa-teramo.inaf.it}

\begin{abstract}
We investigate the distribution of stars along the Horizontal Branch of the Galactic globular cluster 
NGC~1851, to shed light on the progeny of the two distinct Subgiant Branch 
populations harbored by this cluster. On the basis of detailed synthetic Horizontal Branch 
modelling, we conclude that the two subpopulations 
are distributed in different regions of the observed Horizontal Branch: 
the evolved stars belonging to the bright Subgiant branch component are confined 
in the red portion of the observed sequence, whereas the ones belonging 
to the faint Subgiant branch component are distributed from the blue to the red, populating also 
the RR Lyrae instability strip. Our simulations strongly suggest that it is not possible 
to reproduce the observations assuming that the two subpopulations lose the 
same amount of mass along the Red Giant Branch. We warmly encourage empirical estimates 
of mass loss rates in Red Giant stars belonging to this cluster.

\end{abstract}


\section{Introduction}\label{introduction}

A recent $HST$~ACS photometry of the Galactic globular cluster NGC1851 (Milone et al.~2008 -- hereafter M08) 
has disclosed the existence of two distinct Subgiant branches (SGBs) in its Color-Magnitude-Diagram (CMD). 
With this discovery, NGC1851 joins NGC2808 and $\omega$~Centauri in the group of 
globular clusters with a clear photometric signature of multiple stellar populations. 
M08 estimated an age difference of about 1~Gyr between the two subpopulations, in the 
assumption that they share the same initial [Fe/H] (spectroscopy of a few Red Giant Branch 
stars by Yong \& Grundahl~2008, together with the narrow RGB sequence in the CMD confirm this assumption), 
He mass fraction $Y$ and the same metal mixture. Cassisi et al.~(2008 -- hereafter Paper~I) 
have explored the possibility that one of the two sub-populations 
was born with a different heavy element mixture (hereafter denoted as extreme) 
characterized by strong anticorrelations among the CNONa abundances, with a total 
CNO abundance increased by a factor of 2, compared to the normal 
$\alpha$-enhanced metal distribution of the other component (see Sect.~1 and 4 in Paper~I 
for a brief discussion about this choice). 
Both initial chemical compositions share the same [Fe/H] and $Y$ values. 
If the faint SGB component (hereafter SGBf subpopulation) has formed with the extreme metal mixture, 
it results to be coeval with the bright SGB component (hereafter SGBb subpopulation). 
If the reverse is true, the SGBb (extreme) 
subpopulation has to be about 2~Gyr younger than the SGBf (normal) one.
Following the considerations in Paper~I and M08 about the width of Main Sequence and Red Giant Branch (RGB), 
the slope of the Horizontal Branch (HB) in the cluster CMD, plus 
the [Fe/H] estimates by Yong and Grundahl~(2008), one can  
also conclude that appreciable variations of the initial He abundance in the two subpopulations are ruled out.

In this paper we have gone a step further, focusing our attention on the HB. 
Based on the similarity of the number ratio of the 
SGBb to the SGBf components, with the ratio of stars at the red of the instability strip to stars 
at the blue side of the strip, M08 hypothesized that the progeny of the two SGBs occupy separate locations 
along the HB. Here we have addressed this issue in much more detail, analyzing the distribution of 
stars along the HB by means of synthetic HB models, 
to determine whether and for which choice of RGB mass loss, the progeny of the SGBb and SGBf subpopulations 
is able to reproduce the observed HB stellar distribution. We will also readdress the issue, from the point 
of view of HB modeling, 
of whether the extreme metal mixture of Paper~I is compatible with the observed stellar distribution along the 
HB. Section~\ref{models} describes briefly the HB evolutionary tracks and the HB synthetic modeling, while 
Sect.~\ref{results} presents and discusses the results of our analysis.

\section{Synthetic HB modeling}\label{models}

We employed three grids of 39 HB evolutionary tracks each (covering the range between 0.47 and 0.80$M_{\odot}$) 
all with [Fe/H]=$-$1.31 (appropriate for 
NGC1851) calculated by Pietrinferni et al.~(2006) for the normal $\alpha$-enhanced metal mixture, 
and from Paper~I for the extreme mixture, respectively. For the latter models, two different He abundances 
have been adopted, namely: $Y$=0.248 and $Y$=0.280. The models 
have been computed using initial He-core masses 
derived from the evolution of a progenitor with an age of about 12-13~Gyr at the RGB tip. 
All tracks have been normalized to the same number (450 from the start  
to the end of central He-burning) of equivalent points \footnote{See Pietrinferni et al.~(2004) for a discussion on this issue}; 
this simplifies the interpolation to obtain tracks for masses not included in the grid. 

\begin{figure}[t]
\plotone{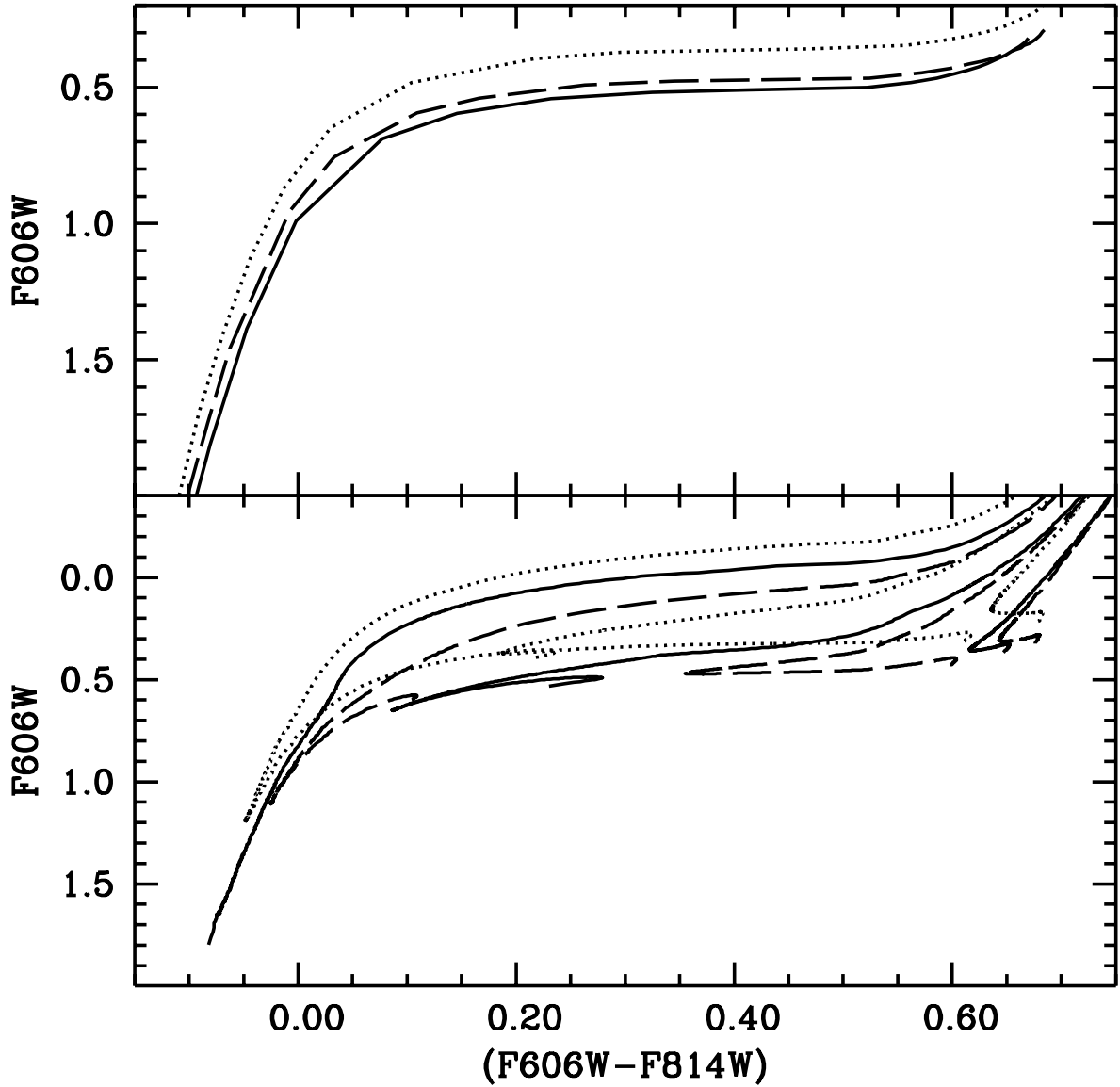}
\caption{Upper panel: ZAHB location in the $F606W - (F606W-F814W)$ plane for, respectively, normal $\alpha$-enhanced 
metal mixture with  $Y$=0.248 and [Fe/H]=$-$1.31 (solid line), extreme mixture with $Y$=0.248 and [Fe/H]=$-$1.31 (dashed line), 
extreme mixture with $Y$=0.280 and [Fe/H]=$-$1.31 (dotted line -- see text for details).
Lower panel: HB evolutionary tracks for masses equal to 0.57, 0.61 and 0.72 $M_{\odot}$, respectively, and the same three 
chemical compositions as in the upper panel.}
\label{fig1}
\end{figure}

Figure~\ref{fig1} displays the Zero Age Horizontal Branch (ZAHB) and selected HB tracks 
for the three chemical compositions considered. The ZAHB for the extreme mixture 
with enhanced He is much brighter than the case of the normal $\alpha$-enhanced composition. This makes very difficult the 
coexistence of sub-populations with these two compositions along the observed HB. The extreme population with 
the same He abundance of the normal one has a much closer ZAHB brightness, a consequence of similar He-core masses 
and surface He abundances of their progenitors at the He-flash. Comparisons of HB tracks for selected 
values of mass display some interesting features. As already shown in Paper~I, models for the extreme 
compositions have a ZAHB location systematically redder than their counterpart 
with the same mass and a normal composition, but the 
blue loops during the He-burning phase are more extended (for a given mass). The 
effect of these differences in terms of mass distribution along the observed HB can be thoroughly assessed only 
by means of synthetic HB modeling.

Synthetic HBs have been calculated as pioneered by Rood (1973). 
The observed HB is simulated by a distribution of stars with different mass, that has 
to be specified as input parameter, together with the time $t$ since each star has first 
arrived on the HB (and the photometric error, obtained from the photometry). 
It is assumed that stars are being fed onto the HB at a constant rate.
Once the stellar mass and $t$ are specified, a quadratic interpolation in mass among the available 
tracks and a linear interpolation in time along the track determine the location 
of the object on the synthetic HB. 
The large number of points along each of our HB tracks ensures that a linear interpolation in 
time is adequate. The magnitudes of the synthetic star are then perturbed by a random value for the 
photometric error in both $F606W$ and $F814W$, according to a Gaussian distribution with dispersion provided by 
the photometric analysis (the typical value in our case is $\sim$0.01~mag).
Either a Gaussian -- the standard assumption, see, e.g. Rood~(1973), Catelan~(1993), Lee et al.~(1990) -- 
or a uniform mass distribution is assumed for the objects fed onto the HB, with mean value $\overline{M}$ and 
dispersion $\sigma$ as free parameters. This is equivalent to assume that the amount of mass lost 
along the previous RGB phase is stochastic, with a specified unimodal distribution. 

Previous analyses of NGC1851, that considered only one  
stellar population in the cluster, have reached different conclusions about 
the unimodality of the RGB mass loss. Lee, 
Demarque \& Zinn~(1988) were able to reproduce the $B:V:R$ number ratio between 
stars located at the blue side of the 
RR~Lyrae instability strip, within the strip and at its red side, using a Gaussian mass distribution. 
Catelan et al.~(1998) also  
conclude that the HB morphology of the cluster can be reproduced with a unimodal (Gaussian) mass distribution, 
assuming a large 1 $\sigma$ dispersion (0.055 $M_{\odot}$). 
On the other hand, Saviane et al.~(1998) conclude that a bimodal mass loss is needed to reproduce 
both the red and blue tails of the observed HB. 

In our simulations, we consider two separate components coexisting on the HB, 
originated from either the SGBb or the SGBf subpopulations. 
The pairs of values ($\overline{M}$, $\sigma$) are left free to vary between the two subpopulations. 
As a consequence -- once the chemical composition of the SGBf and SGBb subpopulations is fixed -- 
one synthetic HB realization is determined by the two pairs 
of ($\overline{M}$, $\sigma$) values chosen for the two components.
The reference HB photometry is the $HST$/ACS ($F606W,F814W$) data by M08. Due to the  
small number of exposures, M08 could not determine appropriate mean magnitudes for the RR Lyrae population;  
therefore only stars detected at the blue (143 objects) and red (242 objects) side of the instability strip are taken 
into account in our comparisons, plus the values of the $B:V:R$ ratios taken from independent data. 
We assume as reference the values 
$30:10:60$, obtained from the number counts provided by Catelan et al.~(1998), extracted 
from Walker~(1992) photometry. They are consistent 
with analogous estimates by Lee et al.~(1988), Walker~(1998) and Saviane et al.~(1998). 
Poisson statistics introduce uncertainties by, respectively, 
$\pm$5\%, $\pm$3\% and $\pm$8\% in the $B:V:R$ values. 
These ratios also agree with the $B:R$ ratio of $(37\pm9):(63\pm7)$ 
determined by M08 from their CMD.

To be considered a match to the observations, a synthetic HB model is required to satisfy the following 
constraints: (i) the empirical values of the ratios $B:V:R$; 
(ii) the $F606W$ and $F814W$ magnitude distribution of the HB stars at the blue 
and red side of the instability strip; (iii) the number ratio of the progeny of the SGBb to the progeny of the SGBf 
subpopulations has to satisfy the 55:45 ratio observed along the SGB. In case the two subpopulations  
share the same age (extreme SGBf component and normal SGBb component) evolutionary times along SGB and RGB 
are such that the 55:45 ratio is conserved also at the beginning of the HB phase. If the extreme component is 
2~Gyr younger (belongs to the SGBb) or both components share the normal composition, population 
ratios at the beginning of the HB phase are altered by only a few percent, and this 
is taken into account in our simulations.

In practice, we have first corrected the observed magnitudes and colors of HB stars by the values of reddening 
($E(F606W-F814W)$=0.04) and distance modulus ($(m-M)_{F606W}$=15.52) determined in Paper~I. 
The resulting magnitudes have then been used for the comparison with our synthetic models.
After selecting the chemical compositions of the two subpopulations -- for each SGB component we considered 
alternatively either a normal or an extreme metal mixture -- we calibrated $\overline{M}$ and 
$\sigma$ (adopting either a Gaussian or a 
uniform mass distribution) for their progenies 
by reproducing the observed $B:V:R$ values, with the additional constraint posed by the 
ratio between the SGBb and SGBf components. All synthetic stars falling in the gap between the red and 
blue HB stars selected by M08 are considered to be RR Lyrae variables. 
We wish to stress that we did not make any a priori choice of where the SGBb and SGBf 
progenies should be located along the observed HB. 

If the $B:V:R$ constraint was satisfied, we finally compared the magnitude distributions of the synthetic 
red and blue HB, with their observational counterparts, by means of a 
Kolmogorov-Smirnov (hereafter KS) test, as applied by Salaris et al.~(2007) 
to the analysis of the HB of 47~Tuc.  
The synthetic HB is considered to be consistent with observations if the KS-test gives a probability 
below 95\% that the observed and synthetic magnitude distributions (in both $F606W$ and $F814W$) 
for both red and blue HB stars are different. 
The number of HB stars in the simulations is typically 20 times larger than the observed value 
in M08 photometry. In this way we minimize, in the synthetic HB model, 
the effect of statistical fluctuations in the number of objects at a given magnitude and color. 

\section{Results and discussion}\label{results}

\begin{figure}[t]
\plotone{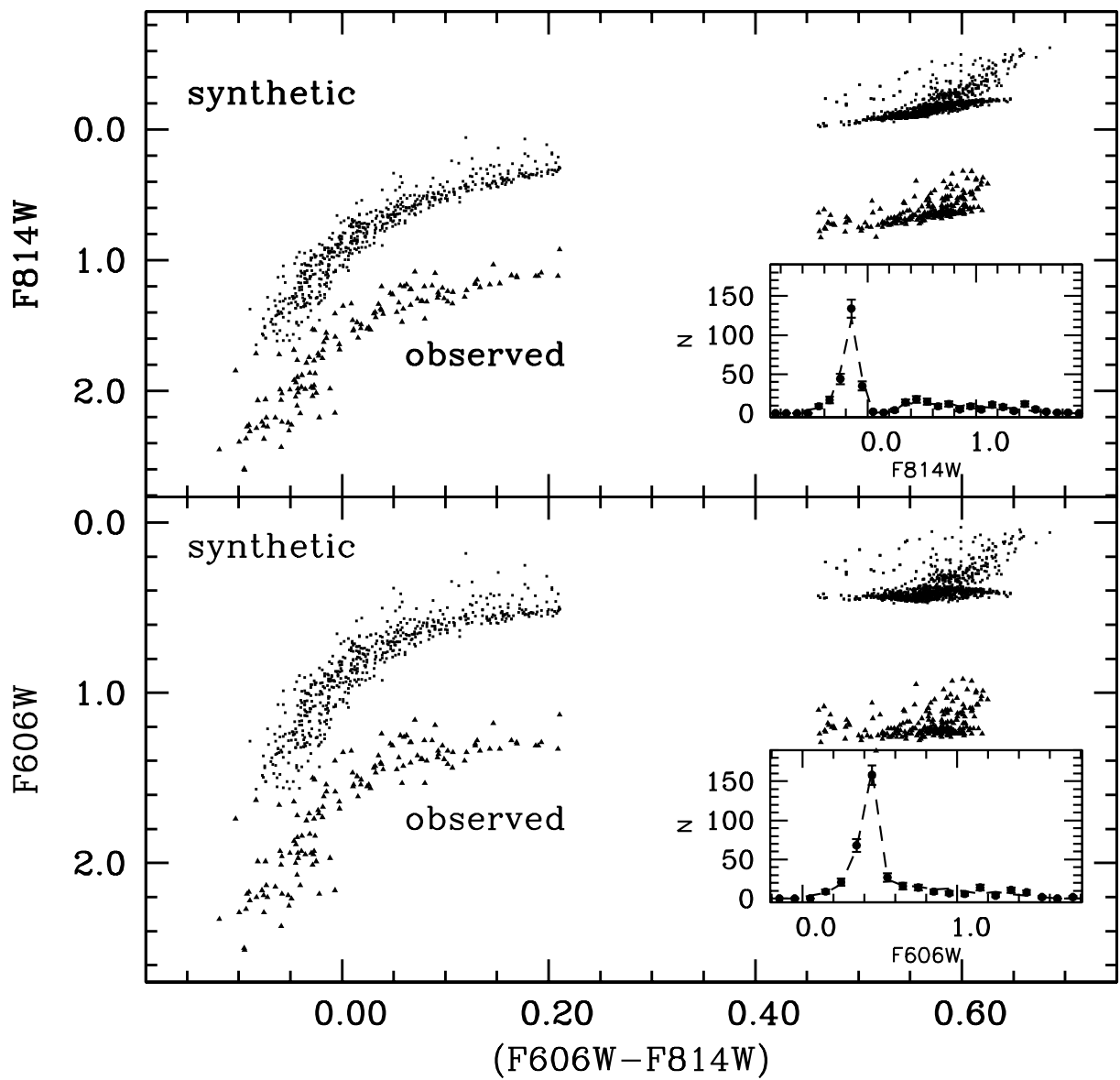}
\caption{Qualitative comparison of the synthetic (Model~2) and observed CMDs of the HB of NGC1851. Observed colors have been 
corrected for the effect of reddening, whereas the $F606W$ and $F814W$ magnitudes have been shifted arbitrarily for 
the sake of clarity. A comparison of star counts along the HB is also shown. Model (dashed lines) counts have been 
rescaled to the same number of observed HB objects. The drop to zero around $F814W$=0 corresponds 
to the gap at the RR Lyrae instability strip in the $F814W$-($F606W-F814W$) CMD.
}
\label{fig2}
\end{figure}

We obtain only one solution for the case 
where both subpopulations have a normal metal mixture (Model~1) as assumed by M08, and 
one solution for the case where one the two subpopulations is characterized by the extreme metal mixture 
(Model~2), i.e. the scenario proposed in Paper~I. 
Figure~\ref{fig2} shows a qualitative comparison between the synthetic HB of Model~2 and the observed one, 
plus the synthetic star counts of the same simulation  
(normalized to the observed number of HB stars in M08 photometry) against the 
observational counterpart. A similar agreement is achieved also for Model 1.
The relevant parameters of the two solutions are summarized in Table~\ref{HBdat}.

Some important conclusions can be drawn from our synthetic HB analysis.
First of all, the progeny of the SGBb subpopulation must be restricted to the 
red part of the HB, whereas the progeny of the 
SGBf component has to be distributed from the 
blue to the red, including the whole instability strip, otherwise the 
KS-test and $B:V:R$ ratios cannot be simultaneously satisfied.
As a consequence, it is not possible to reproduce the observed 
HB by assuming that RGB stars belonging to the 
two subpopulations lose on average the same amount of mass. 
In Model~1 the mass evolving at the RGB tip (in absence of mass loss) for the 
SGBf component is 0.023$M_{\odot}$ smaller than for the SGBb 
one (following the age estimates in Paper~I), whereas the HB modeling 
requires an upper mass limit smaller by more than 
0.035$M_{\odot}$ for its progeny. 
In Model~2 the mass evolving at the RGB tip (in absence of mass loss) 
for the SGBf sub-population 
is only 0.003$M_{\odot}$ smaller than for the SGBb one, 
whereas the HB modeling requires a maximum mass 0.054$M_{\odot}$ smaller. 
The observed $B:V:R$ ratios and the KS-analysis impose two further 
constraints on the model parameters. The mass spread for the red HB 
stars (the progeny of the SGBb sub-population) has to be small, $\sigma < 0.005 M_{\odot}$.
On the other hand, a uniform distribution spanning a large mass range 
is required to reproduce the stellar distributions along the blue part of 
the observed HB (due to the progeny of 
the SGBf subpopulation). Although $B:V:R$ ratios can still be reproduced in 
Model~1 and Model~2 with a Gaussian 
distribution for the SGBf component 
($\overline{M}=0.605M_\odot$ and $\sigma=0.01M_\odot$ for an SGBf population 
with normal metal mixture, and $\overline{M}=0.585M_\odot$ and $\sigma=0.01M_\odot$ 
for an extreme metal mixture), the resulting magnitude distributions are at odds with observations. 

The results in Table~\ref{HBdat} constrain also the relative ages of the SGBb 
and SGBf subpopulations for the scenario of Paper~I.  
In Model~2 it is the SGBf population that is characterized by the extreme composition; this  
corresponds to the case discussed in Paper~I, where SGBb and SGBf 
subpopulations share the same age. A SGBb subpopulation with 
the extreme mixture cannot be accommodated on the HB while at the same time satisfying 
all empirical constraints described above. 
 
We also considered the case of a 50:50 ratio 
between the SGBb and SGBf subpopulations, that is still allowed 
-- within the errors on the measured ratio -- by M08 data. 
Synthetic models with the  
parameter choices of Model~1 and ~2 still match the 
observations (the predicted $B:V:R$ ratios are altered within the 
errors on the reference values). The mass loss necessary to 
reproduce the HB must still differ between the two sub-populations. 
It is also obvious that in this 50:50 scenario an extreme subpopulation can be 
either the SGBf or the SGBb one (therefore being 
either coeval or 2~Gyr younger than the normal component) but its progeny 
has still to populate the extended region from the blue tail of the HB to the red side, 
otherwise the KS-test is not satisfied. 

We have also tested the possibility of having a subpopulation with a mild 
He-enhanced ($Y$=0.28) extreme mixture, but no match to the magnitude distribution 
along the observed HB in the CMD of M08 can ever be achieved.

A different efficiency of mass loss in stars belonging to the same cluster seems difficult to justify, 
especially if both subpopulations are assumed to share the same metal mixture, but given 
the lack of an established theory for the RGB mass loss, we can only use the constraints posed by the HB modeling. 
If different populations in the same cluster lose different amounts  
of mass, the second parameter phenomenon in Galactic globulars may well be at least partly due simply to 
different mass loss efficiencies in different clusters.
These conclusions can in principle change if one hypothesizes a multimodal mass loss, or unimodal probability distributions  
more complex than the standard Gaussian or uniform cases. 
But more free parameters will have to be included and the 
predictive power of synthetic HB modeling would be greatly weakened. 
Overall, our analysis points out that the RGB mass loss in NGC1851 is not simple. 
Either differential mass loss processes are efficient in stars in the same 
cluster, or much more complicated probability distributions for the RGB mass loss may 
have to be employed.  
Empirical determinations of mass loss rates in NGC1851 stars 
(see, e.g., the results by Origlia et al.~2007 for 47~Tuc) are badly needed.  
On this issue, we also wish to note Caloi \& D'Antona~(2008) recent suggestion that a dispersion 
in the initial Helium content among the subpopulations within a single cluster can produce 
the observed HB morphologies, without 
invoking a large dispersion in  the RGB mass loss, or a different mass loss efficiency among the various 
components. In case of NGC~1851 this scenario seems to be disfavored, raising the intriguing possibility that 
different processes affecting the early chemical enrichment and RGB mass loss are at work in different clusters.

Before closing we mention an additional test, involving the cluster pulsators. 
We did not use the constraint posed by their period distribution in our main analysis, because we verified that recent 
theoretical pulsational models of RR Lyrae stars (Di Criscienzo, Marconi \& Caputo~2004 -- see also their discussion about 
uncertainties on the strip boundaries due to the value of the mixing length parameter) predict an instability strip 
for NGC1851 too red by $\sim$ 0.03-0.04 mag in $(F606W-F814W)$ compared to M08 data. 
We have made however the following test, considering the periods determined by Walker~(1998) for 29 cluster RR Lyraes 
(all first overtone pulsators have been fundamentalized by adding 0.13 to the logarithm of their periods in days).
For all synthetic objects of Model~1 and 2 falling in the observed (not theoretical) RR Lyrae gap we determined the pulsation 
period from the fundamental pulsation equation by Di Criscienzo et al.~(2004 -- their Equation~1). 
The period distribution of the synthetic RR Lyrae stars has been then compared to the observed 
one by means of a KS-test. Interestingly, we found that 
Model~1 gives a period distribution inconsistent with observations with a probability larger than 95\%, whereas in case 
of Model~2 this probability is well below the 95\% threshold, and we consider this model to have periods statistically 
in agreement with observations.
  
\acknowledgments{We warmly thank A. Sarajedini for allowing us the use - in this paper as well as in Paper I - of his 
photometric data, and G. Piotto for reading a preliminary draft of this paper and useful suggestions.}



\clearpage
\begin{deluxetable}{lllll}
\tablewidth{0pt}
\tablecaption{Properties of the synthetic models that match the observed HB of NGC~1851\label{HBdat}}
\tablehead{\colhead{Subpopulation} & \colhead{Mixture}  & HB coverage & \colhead{Mass distribution} 
& \colhead{$B:V:R$}}
\startdata
\multicolumn{5}{c}{Model 1}\\
\hline
SGBf    & normal  &  $B+V+R$  & Uniform,   0.570 to 0.625    $M_{\odot}$  & 31:11:58\\
SGBb    & normal  &  $R$      & Gaussian,  $\overline{M}$=0.66  $M_{\odot}$ ($\sigma <$ 0.005 $M_{\odot}$)  &         \\
\hline
\multicolumn{5}{c}{Model 2}\\
\hline
SGBf    & extreme &  $B+V+R$  & Uniform,  0.554 to 0.606    $M_{\odot}$   & 30:10:60\\
SGBb    & normal  &  $R$      & Gaussian, $\overline{M}$=0.66  $M_{\odot}$ ($\sigma <$ 0.005 $M_{\odot}$) &         \\
\enddata
\end{deluxetable}

\clearpage

\clearpage

\end{document}